\documentclass[12pt]{iopart}

\usepackage{graphicx}
\begin{document}

\title[V. Dexheimer et al.]{Neutron Stars in a Chiral Model with Finite Temperature}

\author{V. Dexheimer, S. Schramm, H. Stoecker}

\address{FIAS, Johann Wolfgang Goethe University,
Max-von-Laue-Str. 1,
60438, Frankfurt am Main,
Germany}
\ead{dexheimer@th.physik.uni-frankfurt.de}
\begin{abstract}
Neutron star matter is investigated in a hadronic chiral model
approach using the lowest flavor-SU(3) multiplets for baryons and
mesons. The parameters are determined to yield consistent results
for saturated nuclear matter as well as for finite nuclei. The
influence of baryonic resonances is discussed. The global
properties of a neutron star such as its mass and radius are
determined. Proto-neutron star properties are studied by taking
into account trapped neutrinos, temperature and entropy effects.

\end{abstract}

\maketitle

\section{Introduction}

Although the early stages of the life of a neutron star, called
proto-neutron star, last for just a few seconds, their properties
have consequences that dictate the properties of the neutron star
many years after its formation. For example, the baryon number of
a proto-neutron star is extremely important because it is a limit
for the baryon mass of the neutron star it will form, since it is
known that when the maximum mass predicted by the Einstein's
equations of General Relativity is exceeded the star will collapse
into a black hole \cite{hole}. During the life of the
proto-neutron star, all the electron neutrinos trapped during the
supernova explosion escape carrying thermal energy out of the
stellar system so that the temperature of the star decreases very
quickly during this process. Luckily the time the changes in the
composition of the star takes to occur is more than three orders
of magnitude bigger than the dynamical timescale of the
readjustment of pressure and gravity forces, so that the system
can still be treated in a quasistatic approximation.

An efficient way to describe infinite nuclear matter with hadronic
degrees of freedom is through the use of effective theories
because they allow us to work in a specific energy scale,
practically ignoring other degrees of freedom of the system. Since
not much is known about matter at very high densities, the results
are first compared to nuclear matter properties in order to
constrain the values of the coupling constants of the model and,
as a second step, the macroscopic properties of the stars (such as
mass and radius) are used to select the most acceptable models.

The matter inside neutron stars has densities up to several times
the nuclear saturation density; at this point the scalar
condensates responsible for chiral symmetry breaking of vacuum are
strongly reduced and chirality is largely restored therefore.
based on this feature, we conclude that it is important to have a
chiral invariant lagrangian density \cite{ch}. it is also
important to analyze in which way chiral symmetry restoration
occurs.

\section{The Chiral Model}

Because of the hight densities nuclear matter inside neutron stars
can reach it is possible to find not only nucleons, but also
hyperons ($\Lambda$, $\Sigma$, $\Xi$) and resonances ($\Delta$,
$\Sigma^*$, $\Xi^*$, $\Omega$) in the core of the star. This
system can be described by a nonlinear chiral model in the
mean-field approximation. Besides the kinetic energy part, the
Lagrangian density contains terms that describe the interaction
between baryons and scalar mesons and between baryon and vector
mesons, self interactions of scalar and vector mesons and a term
that breaks chiral symmetry explicitly, which is responsible for
the pseudo-scalar mesons masses. Besides the usual mesons
($\sigma$, $\delta$, $\zeta$, $\omega$, $\phi$, $\rho$), the
dilaton field ($\chi$), acting as the effective gluon condensate,
is also included. Electrons and muons are considered for charge
neutrality. Therefore, the Lagrangian density of our model is:
\begin{eqnarray}
&L_{MFT}=L_{Kin}+L_{Bscal}+L_{Bvec}+L_{scal}+L_{vec}+L_{SB},&
\end{eqnarray}
\begin{eqnarray}
&L_{Bscal}+L_{Bvec}=-\sum_i \bar{\psi_i}[g_{i\omega}\gamma_0\omega+g_{i\phi}\gamma_0\phi+g_{i\rho}\gamma_0\tau_3\rho+m_i^*]\psi_i,&
\end{eqnarray}
\begin{eqnarray}
&L_{vec}=-\frac{1}{2}(m_\omega^2\omega^2+m_\rho^2\rho^2)\frac{\chi^2}{\chi_o^2}-\frac{1}{2}m_\phi^2\phi^2
\frac{\chi^2}{\chi_o^2}&\nonumber\\&-g_4\left[\omega^4+6\rho^2\omega^2+\rho^4+2\phi^4\right],&
\end{eqnarray}
\begin{eqnarray}
&L_{scal}=\frac{1}{2}k_0\chi^2(\sigma^2+\zeta^2+\delta^2)-k_1(\sigma^2+\zeta^2+\delta^2)^2&\nonumber\\&-k_2\left(\frac{\sigma^4}{2}+\frac{\delta^4}{2}
+3\sigma^2\delta^2+\zeta^4\right)-k_3\chi(\sigma^2-\delta^2)\zeta&\nonumber\\&+k_4\chi^4+\frac{1}{4}\chi^4\ln{\frac{\chi^4}{\chi_0^4}}-eps\ \ \chi^4\ln{\frac{(\sigma^2-\delta^2)\zeta}{\sigma_0^2\delta_0}},&
\end{eqnarray}
\begin{eqnarray}
&L_{SB}=\left(\frac{\chi}{\chi_0}\right)^2\Bigg[m_\pi^2 f_\pi\sigma+\left(\sqrt{2}m_k^ 2f_k-\frac{1}{\sqrt{2}}m_\pi^ 2 f_\pi\right)\zeta\Bigg].&
\end{eqnarray}

The baryon masses are entirely generated by the scalar fields
except for a small explicit mass term $\delta$m. The effective
masses decrease at hight densities, since the scalar fields tend
to zero in this limit; so, the effective masses reproduce the
measured baryon masses at low densities:
\begin{eqnarray}
&m^*=g_{i\sigma}\sigma+g_{i\delta}\delta+g_{i\zeta}\zeta+g_{i\chi}\chi+\delta m.&
\end{eqnarray}

\section{Results}

\begin{figure}[!pt]
\begin{center}
\includegraphics[width=0.5\textwidth, clip,trim= 0 0 0 0]{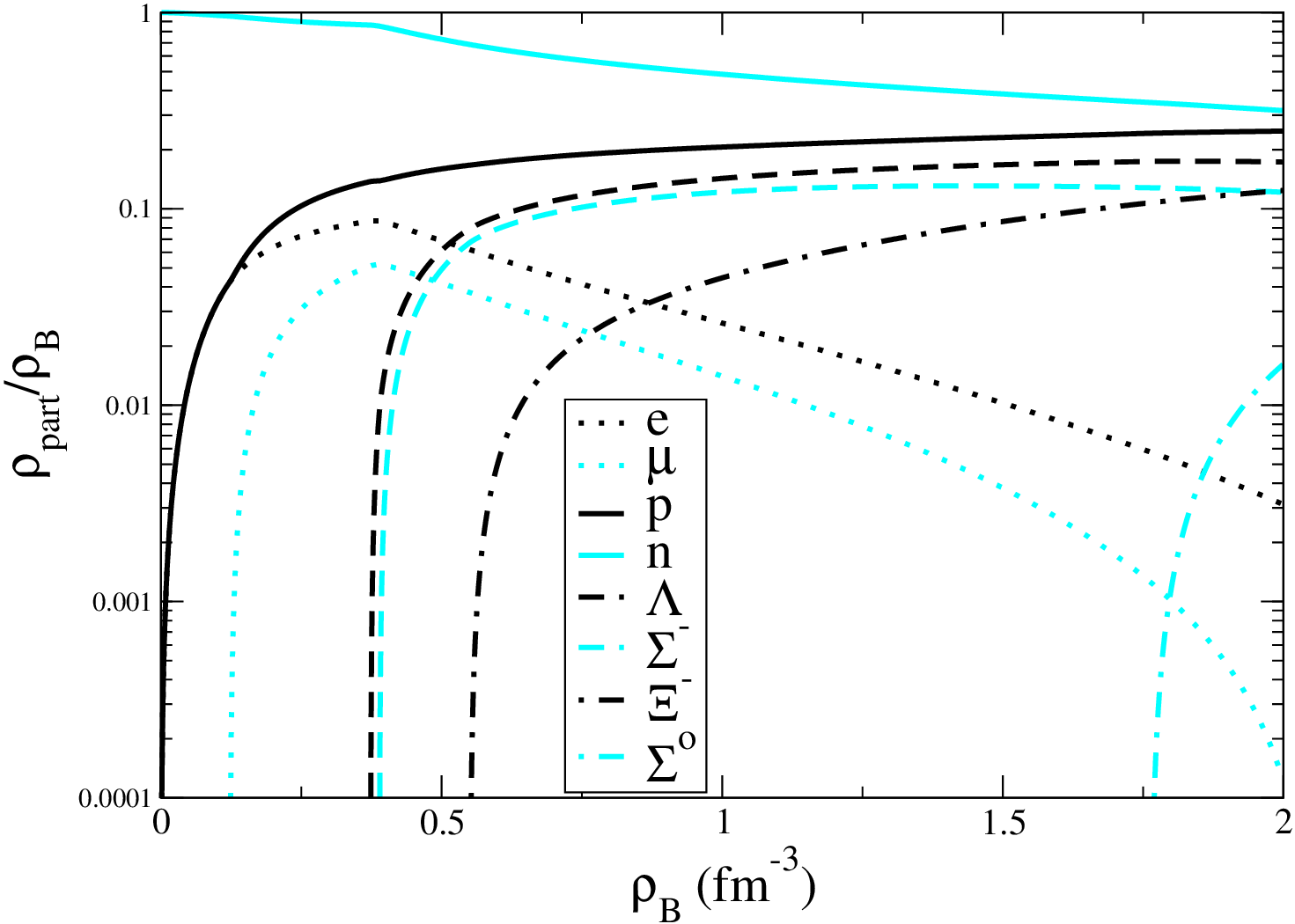}
\end{center}
\caption{The composition of neutron star matter with hyperons}
\label{pop}
\end{figure}

\begin{figure}[!pt]
\begin{center}
\includegraphics[width=0.5\textwidth, clip,trim= 0 0 0 1]{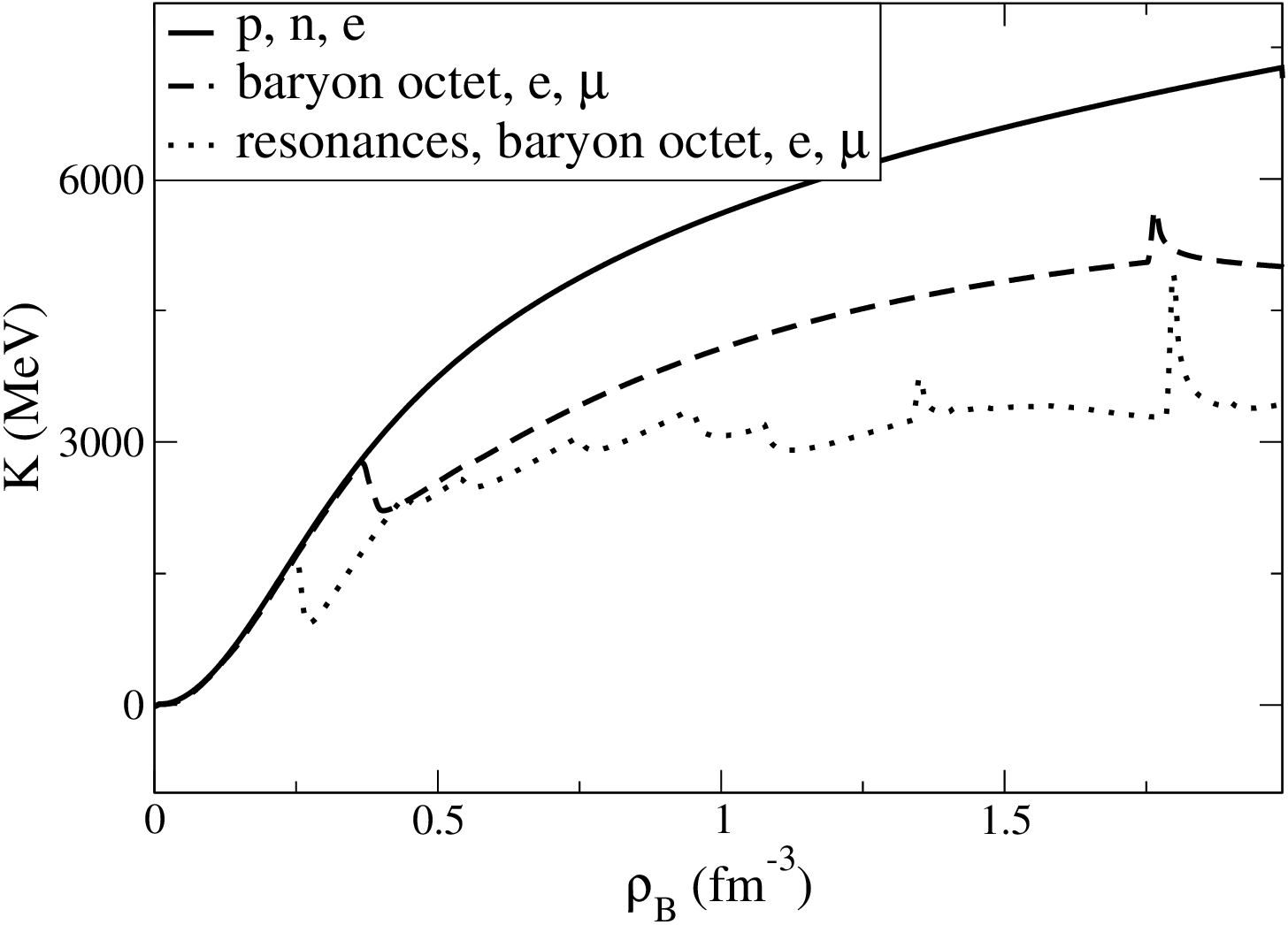}
\end{center}
\caption{Compressibility function \emph{versus} density} \label{k}
\end{figure}

At small densities, the star contains only neutrons. When density
increases towards the center of the star, other particles appear.
First, protons and electrons are populated at the same rate in
order to keep charge neutrality. At higher densities, the charge
neutrality must be achieved taking into account the presence of
hyperons ($\Lambda$, $\Sigma$ and $\Xi$) in the system (figure
\ref{pop}). The appearance of these particles can also be seen in
the compressibility function plot (figure \ref{k}), first
calculated in \cite{eu}. The compressibility function curve
exhibits a bump at each density when a new hyperon species is
populated; that is because compressibility, being a second order
derivative, is a measure of the rate of increase of energy as a
function of density:
\begin{eqnarray}
\label{kdpdrho}K=9\frac{dP}{d\rho}=9\rho\frac{d^2\epsilon}{d\rho^2}.
\end{eqnarray}

\begin{figure}[!pt]
\begin{center}
\includegraphics[width=0.5\textwidth, clip,trim= 0 0 0 0]{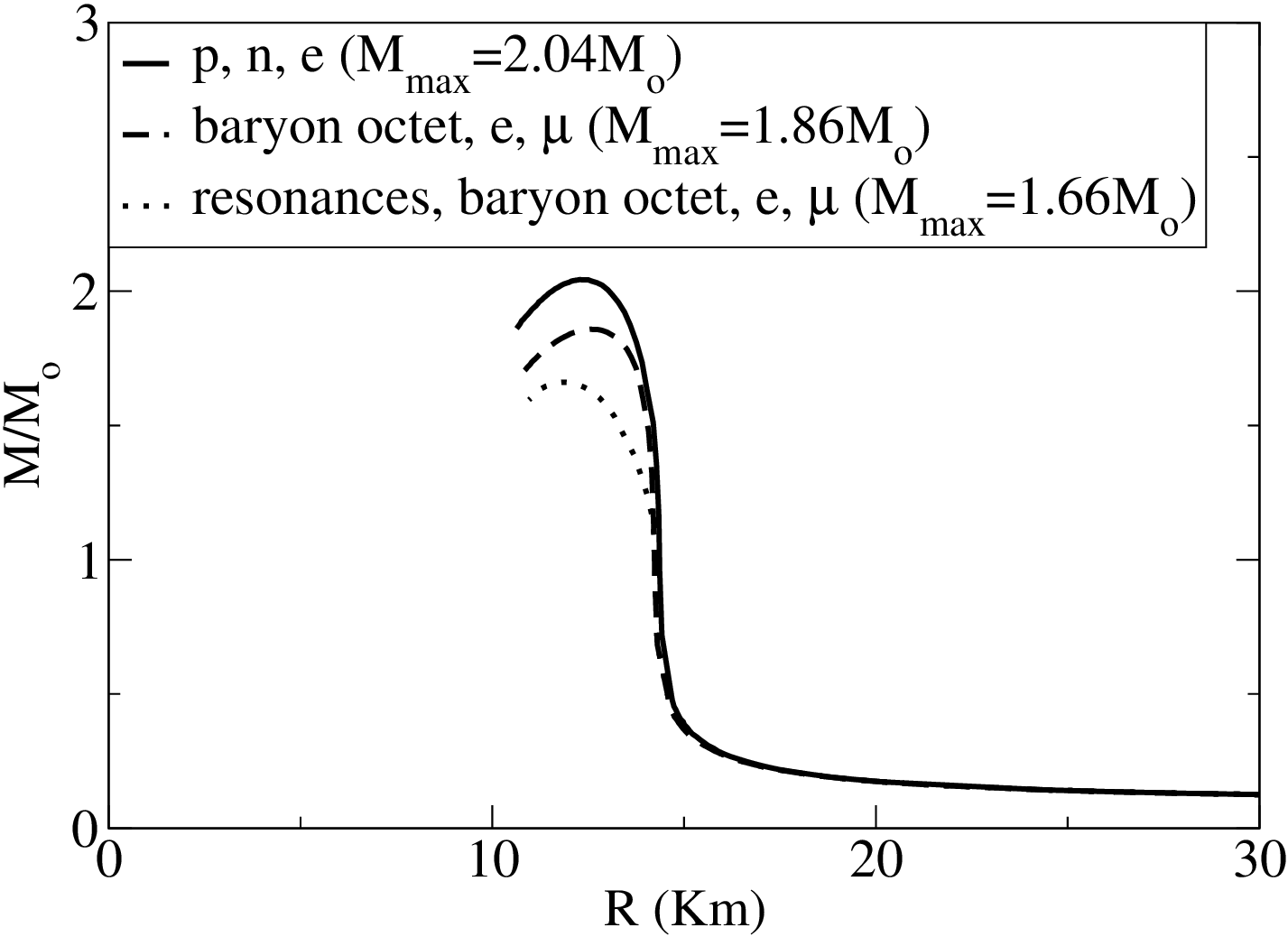}
\end{center}
\caption{Star mass versus radius for different constituents}
\label{mass}
\end{figure}

The negative values the compressibility function assumes for very
small densities are a consequence of the liquid-gas phase
transition. It can also be seen that the mass that the star can
hold against gravitational collapse is larger for systems composed
by higher compressibility nuclear matter, as shown in figure
\ref{mass} (here, besides the model applied for the interior of
the star, an inner crust, an outer crust  and an atmosphere are
also considered). In comparison with a system composed of nucleons
and electrons solely, the maximum star mass predicted when muons
and the whole baryon octet are included is smaller because of the
new degrees of freedom available. The effect of the resonances is
the same, but in this case the star mass predicted is so small
that it can not describe the very massive stars that have been
observed lately.

For proto-neutron stars, the maximum mass increases with
temperature up to a certain limit (40 MeV). After that the layer
formed by electron-positron pairs that surround the star becomes
too big, which requires a more careful treatment.

\begin{figure}[!pt]
\begin{center}
\includegraphics[width=0.5\textwidth, clip,trim= 0 0 0 1]{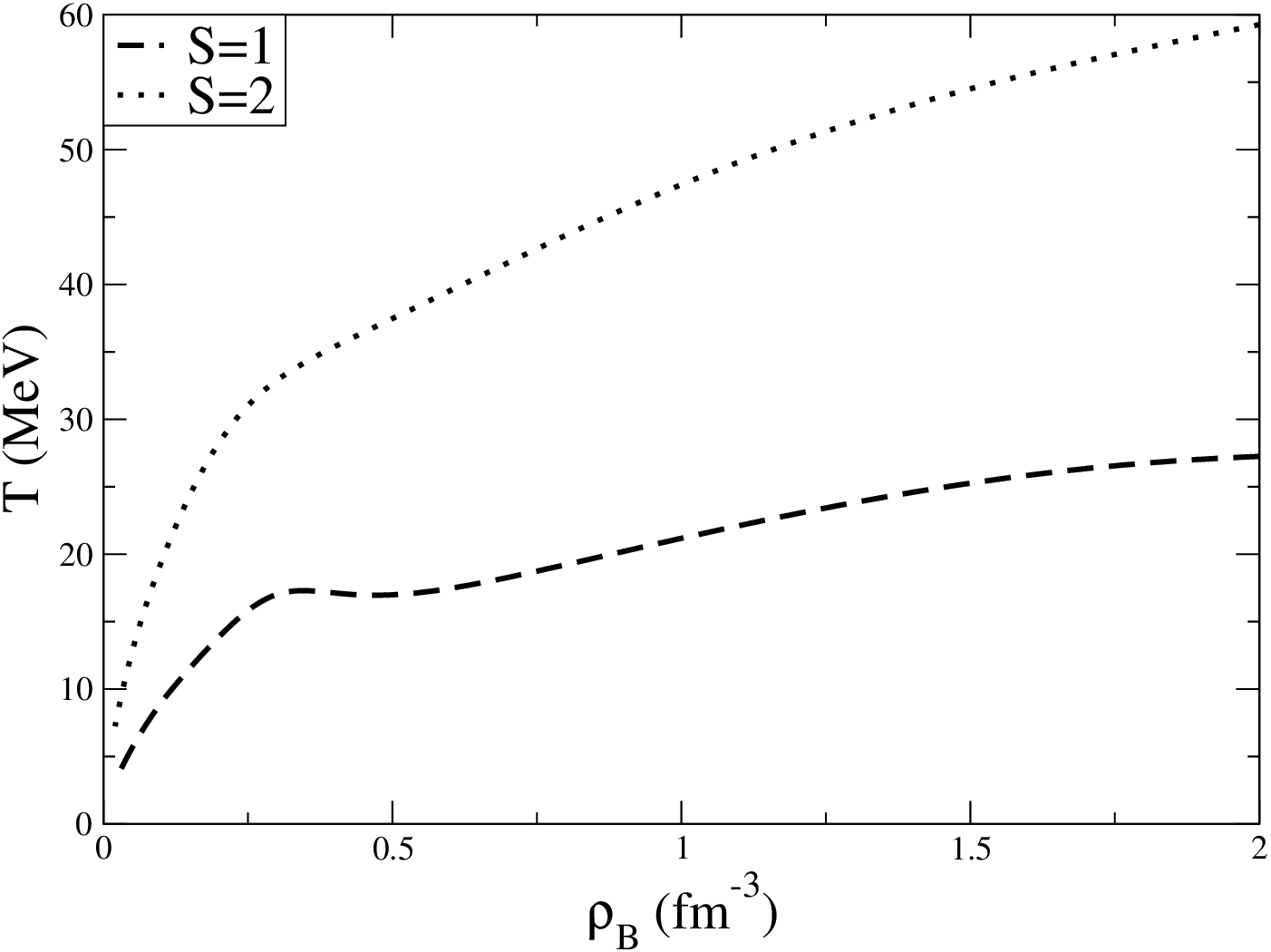}
\end{center}
\caption{Temperature versus density for different entropies.}
\label{S}
\end{figure}

\begin{figure}[!pt]
\begin{center}
\includegraphics[width=0.5\textwidth, clip,trim= 0 0 0 0]{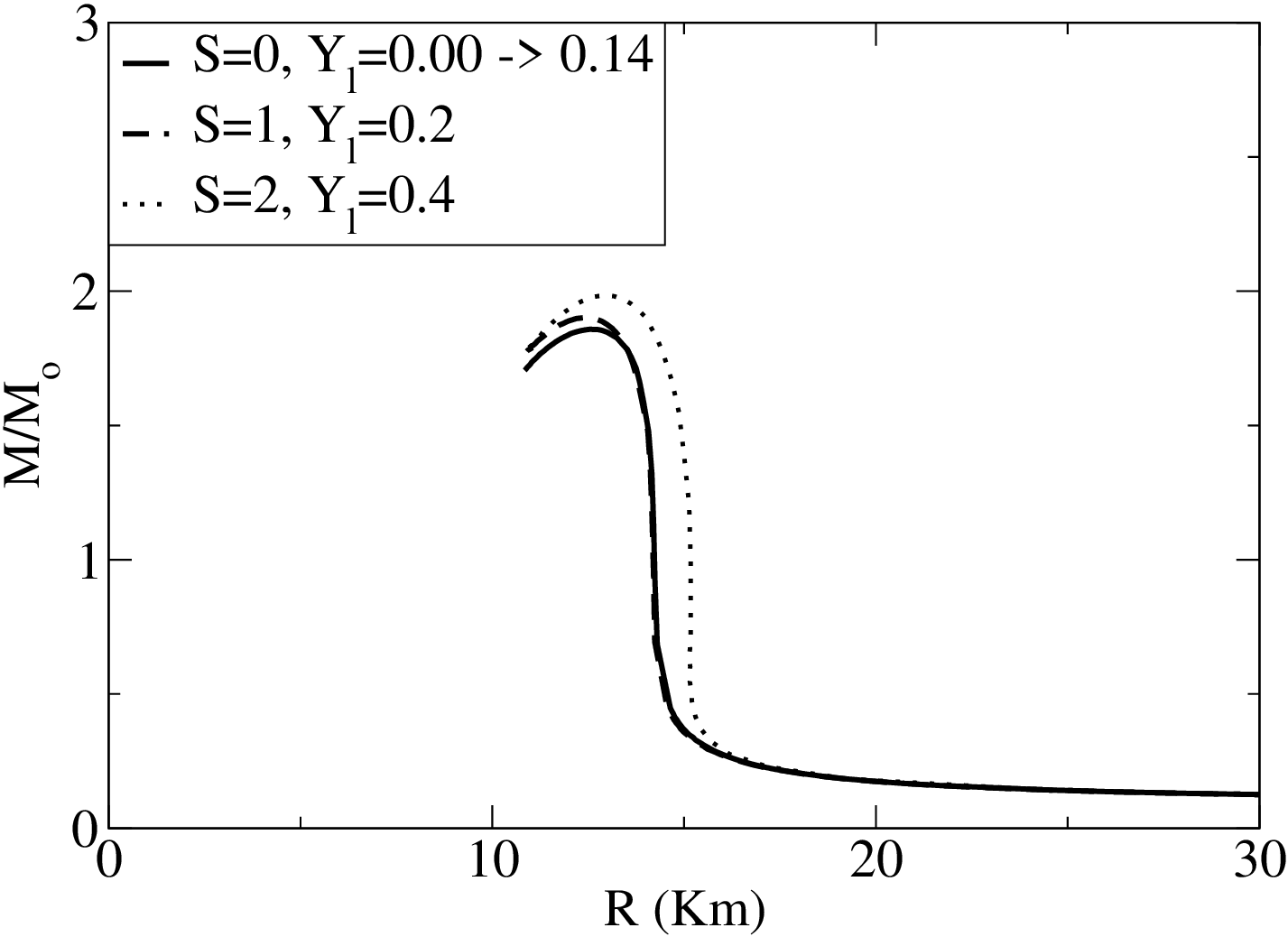}
\end{center}
\caption{Star mass versus radius for different entropies and lepton numbers}
\label{Ms}
\end{figure}

Up to this point of our analysis, the whole star is considered to
be at the same constant temperature, what is unrealistic
since it is known that the stars are hotter in the center. A more
natural approach \cite{ent} would be to consider a star with
constant entropy (figure \ref{S}). In this case, in an early stage
(S/baryon=2) the temperature can reach 60 MeV and in a second
stage (S/baryon=1) the temperature can reach 30 MeV in the center,
but at all stages the star has got a low temperature on the
border. During the evolution of the proto-neutron star, both
temperature and entropy decrease as the neutrinos leave the star,
so the lepton number, defined as
\begin{eqnarray}
&Y_l=\frac{\rho_e+\rho_{\nu_e}}{\rho_B},&
\end{eqnarray}
can be expected to decrease as well. This feature is shown in
figure \ref{Ms}, where it can be seen that the maximum mass of the
star decreases with time. There are two reasons for that: the
first is that higher entropy implies higher temperature at any
density and the second is that higher lepton number implies more
negative charge, and in this case there is no need of many
hyperons in order to achieve charge netruality (and the presence
of hyperons tend to reduce the maximum mass of the star). Even for
the SU(2) case, which contains no hyperons, the effect is the
same. This fact demonstrates the relevance of thermal effects in
this model.

\section{Conclusion}

The chosen set of parameters allows us to model a massive neutron
star which contains the complete octet of baryons. The maximum
mass predicted by this model is $M=1.86M_{\odot}$ while the
heaviest star observed has a mass of
$M=2.1^{+0.4}_{-0.5}M_{\odot}$\cite{21}.

The description of proto-neutron stars is done by fixing entropy
and lepton number, and allowing higher temperatures in the center
of the star. At a first moment, right after the supernova
explosion, approximately described by $S=2$ and $Y_l=0.4$ the star
contains many neutrinos. After several seconds a considerable
fraction of these neutrinos escapes carrying thermal energy out of
the star; as a consequence, the maximum mass of the star decreases
about $4\%$. In a final stage the cold star without neutrinos has
a maximum mass about $6\%$ smaller than in the initial case. This
shows the influence both temperature and the amount of leptons
have in the binding effects that determine the mass of a
proto-neutron star.

\end{document}